\documentclass[12pt]{article}
\usepackage{amssymb}
\usepackage{epsfig}

\makeatletter
\@addtoreset{equation}{section}
\renewcommand{\theequation}{\thesection.\arabic{equation}}
\renewcommand{\thefootnote}{\fnsymbol{footnote}}
\makeatother

\begin{document}
\newcommand{\p}[1]{(\ref{#1})}
\newcommand {\beq}{\begin{eqnarray}}
\newcommand {\eeq}{\end{eqnarray}}
\newcommand {\non}{\nonumber\\}
\newcommand {\eq}[1]{\label {eq.#1}}
\newcommand {\defeq}{\stackrel{\rm def}{=}}
\newcommand {\gto}{\stackrel{g}{\to}}
\newcommand {\hto}{\stackrel{h}{\to}}
\newcommand {\1}[1]{\frac{1}{#1}}
\newcommand {\2}[1]{\frac{i}{#1}}
\newcommand {\thb}{\bar{\theta}}
\newcommand {\ps}{\psi}
\newcommand {\psb}{\bar{\psi}}
\newcommand {\ph}{\varphi}
\newcommand {\phs}[1]{\varphi^{*#1}}
\newcommand {\sig}{\sigma}
\newcommand {\sigb}{\bar{\sigma}}
\newcommand {\Ph}{\Phi}
\newcommand {\Phd}{\Phi^{\dagger}}
\newcommand {\Sig}{\Sigma}
\newcommand {\Phm}{{\mit\Phi}}
\newcommand {\eps}{\varepsilon}
\newcommand {\del}{\partial}
\newcommand {\dagg}{^{\dagger}}
\newcommand {\pri}{^{\prime}}
\newcommand {\prip}{^{\prime\prime}}
\newcommand {\pripp}{^{\prime\prime\prime}}
\newcommand {\prippp}{^{\prime\prime\prime\prime}}
\newcommand {\pripppp}{^{\prime\prime\prime\prime\prime}}
\newcommand {\delb}{\bar{\partial}}
\newcommand {\zb}{\bar{z}}
\newcommand {\mub}{\bar{\mu}}
\newcommand {\nub}{\bar{\nu}}
\newcommand {\lam}{\lambda}
\newcommand {\lamb}{\bar{\lambda}}
\newcommand {\kap}{\kappa}
\newcommand {\kapb}{\bar{\kappa}}
\newcommand {\xib}{\bar{\xi}}
\newcommand {\ep}{\epsilon}
\newcommand {\epb}{\bar{\epsilon}}
\newcommand {\Ga}{\Gamma}
\newcommand {\rhob}{\bar{\rho}}
\newcommand {\etab}{\bar{\eta}}
\newcommand {\chib}{\bar{\chi}}
\newcommand {\tht}{\tilde{\th}}
\newcommand {\zbasis}[1]{\del/\del z^{#1}}
\newcommand {\zbbasis}[1]{\del/\del \bar{z}^{#1}}
\newcommand {\vecv}{\vec{v}^{\, \prime}}
\newcommand {\vecvd}{\vec{v}^{\, \prime \dagger}}
\newcommand {\vecvs}{\vec{v}^{\, \prime *}}
\newcommand {\alpht}{\tilde{\alpha}}
\newcommand {\xipd}{\xi^{\prime\dagger}}
\newcommand {\pris}{^{\prime *}}
\newcommand {\prid}{^{\prime \dagger}}
\newcommand {\Jto}{\stackrel{J}{\to}}
\newcommand {\vprid}{v^{\prime 2}}
\newcommand {\vpriq}{v^{\prime 4}}
\newcommand {\vt}{\tilde{v}}
\newcommand {\vecvt}{\vec{\tilde{v}}}
\newcommand {\vecpht}{\vec{\tilde{\phi}}}
\newcommand {\pht}{\tilde{\phi}}
\newcommand {\goto}{\stackrel{g_0}{\to}}
\newcommand {\tr}{{\rm tr}\,}
\newcommand {\GC}{G^{\bf C}}
\newcommand {\HC}{H^{\bf C}}
\newcommand{\vs}[1]{\vspace{#1 mm}}
\newcommand{\hs}[1]{\hspace{#1 mm}}
\newcommand{\al}{\alpha}
\newcommand{\be}{\beta}
\newcommand{\Lam}{\Lambda}
\newcommand{\kahler}{K\"ahler }
\newcommand{\con}[1]{{\Gamma^{#1}}}
\newcommand{\sect}[1]{\setcounter{equation}{0}\section{#1}}
\renewcommand{\theequation}{\thesection.\arabic{equation}}

\thispagestyle{empty}
\begin{flushright}
{\tt HIP-2008-08/TH}\\
{\tt KEK-TH-1242}
\end{flushright}
%
%
\begin{center}
{\Large
{\bf 
Influence of $Z'$ boson \\
on top quark spin correlations at the LHC
}}
\\[8mm]
\vspace{3mm}

\normalsize
{\large \bf
  Masato~Arai~$^{a~b}$}
\footnote{\it
masato.arai@gmail.com}
,
{\large \bf
  Nobuchika~Okada~$^{c}$}
\footnote{\it
okadan@post.kek.jp}
,
{\large \bf Karel Smolek~$^{d}$}
\footnote{\it
karel.smolek@utef.cvut.cz
}
\\
and 
\\
{\large \bf
Vladislav \v{S}im\'ak~$^{e}$}
\footnote{\it
simak@fzu.cz
}

{
 $^{a}$ {\it Center for Quantum Spacetime (CQUeST), Sogang University, \\
  Shinsu-dong 1, Mapo-gu, Seoul 121-742, Korea}\footnote{Present address}\\
 $^{b}$ \it  Department of Physics,
             University of Helsinki \\
 and Helsinki Institute of Physics,
 P.O.Box 64, FIN-00014, Finland
         \\
 $^{c}$ \it Theory Division, KEK, Tsukuba, 
        Ibaraki 305-0801, Japan \\
 $^{d}$ \it Institute of Experimental and Applied Physics, \\
        Czech Technical University in Prague, 
        Horsk\'a 3a/22, 128 00 Prague 2, Czech Republic \\
 $^{e}$ \it  Faculty of Nuclear Sciences and Physical Engineering,\\
        Czech Technical University in Prague, 
        B\v{r}ehov\'a 7, 115 19 Prague 1, Czech Republic
}

%
%
%
{\bf Abstract}\\[5mm]
{\parbox{16cm}{
We study top-antitop pair production and 
 top spin correlations in a model with an electrically neutral massive
 gauge boson, $Z^\prime$, at the Large Hadron Collider.
In addition to the Standard Model processes,
 the $Z'$ contributes to the top-antitop pair production process in the 
 $s$-channel.
Choosing a kinematical region of top invariant mass
 around the $Z^\prime$ resonance pole,
 we find sizable deviations of the top-antitop pair production cross
 section and the top spin correlations from those of the Standard Model.
}}
\end{center}
\vfill
\newpage
\setcounter{page}{1}
\setcounter{footnote}{0}
\renewcommand{\thefootnote}{\arabic{footnote}}
%
%
\section{Introduction}
The Standard Model (SM) based on the gauge group $SU(3)\times SU(2)_L\times U(1)_Y$
 is phenomenologically quite successful in
 the description of phenomena around the electroweak scale.
However, it is widely believed that new physics beyond the SM takes
 place above certain energy scale.
In a class of new physics models,
 the SM gauge group is embedded in a larger gauge group.
Such a model often predicts an 
 electrically neutral massive gauge boson, referred as $Z'$.
There are many example models such as the left-right symmetric model 
 \cite{LR} and Grand Unified Theories based on the gauge 
 groups $SO(10)$ \cite{SO10} and $E_6$ \cite{E6} (for a review, see, for
 example, \cite{leike,Rizzo,Langacker}).

At the hadron collider a
 $Z^\prime$ boson could be observed as a resonance
 in the Drell-Yan process.
Current direct search of $Z'$ bosons for several models 
 has been performed by the CDF
 collaboration at the Tevatron in the $e^+e^-$ decay channel with 
 the use of the di-electron invariant mass and angular distributions.
No evidence of a signal has been found and $95\%$ CL
 lower limits of the $Z'$ mass are set to be in the range from 
 650 to 900 GeV \cite{CDF}.
Recently studies about measurement of $Z'$ bosons at the 
 Large Hadron Collider (LHC) has been performed \cite{fuks,petriello,guzzi}.

If a $Z'$ boson mass lies around TeV scale, it could be discovered at
 the LHC.
Once a $Z^\prime$-like resonance is observed, the next task is to
 precisely measure its properties, such as mass, spin, couplings to the
 SM particles etc.
The spin of the $Z'$ boson affects angular distributions and 
 spin configurations for outgoing particles produced by $Z'$ decays.
A good tool to study the spin configuration is 
 a top-antitop quark pair.
Since the top quark with mass in the range of $175$ GeV \cite{Abe} 
 decays electroweakly before hadronization takes place \cite{Bigi},
 a spin polarization of the top-antitop quark pair 
 is directly transferred to its decay products.
Therefore there are significant angular correlations 
 between the top quark spin axis and the direction of motion 
 of the decay products. 

The spin correlations for the hadronic top-antitop pair production 
 process have been extensively studied in the quantum 
 chromodynamics (QCD) \cite{Stelzer, Mahlon-Parke, Bernreuther2}. 
It is found that there is a spin asymmetry between the produced 
 top-antitop pairs, namely, 
 the number of produced top-antitop quark pairs with both spin up or 
 spin down (like pair) is different from the number of pairs 
 with the opposite spin combinations (unlike pair). 
If the top quark is coupled to new physics beyond the SM,
 it could alter the top-antitop spin correlations. 
Therefore, the top-antitop spin correlations can provide 
 useful information to test not only the SM but also 
 a possible new physics at the hadron collider.
The LHC has a big advantage to study the top spin correlations, 
 since it will produce almost 10 millions of top quarks a year. 
In Refs. \cite{AOSS1} and \cite{AOSS2}, 
 effects of the Kaluza-Klein gravitons on the top spin correlations 
 in the brane world models at the LHC were studied and 
 sizable deviations of the top spin correlations from the SM one were
 found\footnote{There are also several studies of effects of new physics 
 on the top spin correlations at 
 $e^+e^-$ collider \cite{ee} and photon collider \cite{gamma}.}.

The purpose of this paper is to investigate effects of the
 $Z^\prime$ boson on the top-antitop pair production and its spin correlations.
In addition to the SM processes, 
 the $Z'$ boson gives rise to a new contribution for the
 top-antitop pair production process in the $s$-channel and
 alters the top-antitop pair production cross
 section and the top spin correlations from the SM ones.
Choosing a kinematical region of top invariant mass
 around the $Z^\prime$ resonance pole,
 we find their sizable deviations from those of the SM. 
 
This paper is organized as follows.
In the next section, we briefly review the top spin correlations.
In section 3, we present a model with the $Z'$ boson.
In section 4, we derive the invariant amplitudes for the polarized 
 top-antitop pair production processes mediated by the $Z^\prime$ 
 boson in the $s$-channel. 
We show the results of our numerical analysis in section 5. 
Section 6 is devoted to conclusions.
Appendices ensemble formulas we use in the calculation.

\section{Spin correlation}
At hadron collider, the top-antitop quark pair is produced 
 through the processes of quark-antiquark pair annihilation 
 and gluon fusion:
\begin{eqnarray}
 &i \rightarrow t+\bar{t}, \,\,\, i=q\bar{q}\,,gg\,.& \label{top1}
\end{eqnarray}
The former is the dominant process at the Tevatron, 
 while the latter is dominant at the LHC. 
The produced top-antitop pairs decay before hadronization takes place. 
The main decay modes in the SM 
 involve leptonic and hadronic modes: 
\begin{eqnarray}
 &t\rightarrow bW^+ \rightarrow bl^+\nu_l\,,bu\bar{d}\,,bc\bar{s},&
 \label{decay}
\end{eqnarray}
where $l=e,\mu,\tau$.
The differential decay rates to a decay product $f=b,l^+, \nu_l,$ etc.  
 at the top quark rest frame can be parameterized as 
\begin{eqnarray}
{1 \over \Gamma}{d \Gamma \over d \cos \theta_f}=
  {1 \over 2}(1 + \kappa_f \cos \theta_f ), 
 \label{decay1}
\end{eqnarray}
where $\Gamma$ is the partial decay width of 
 the respective decay channel and 
 $\theta_f$ is the angle between the 
 top quark polarization
 and the direction of motion of the decay product $f$.
The coefficient $\kappa_f$ called the top spin analyzing power 
 is a constant between $-1$ and $1$.
The ability to distinguish 
 the polarization
 of the top quark evidently increases with $\kappa_f$.
The most powerful spin analyzer is a charged lepton, for which 
 $\kappa_{l^+}=+1$ at tree level \cite{Jezabek}. 
Other values of $\kappa_f$ are
 $\kappa_b = -0.41$ for the $b$-quark 
 and $\kappa_{\nu_l}=-0.31$ for the $\nu_l$, respectively. 
In hadronic decay modes, the role of the charged lepton 
 is replaced by the $\bar{d}$ or $\bar{s}$ quark.

Now we see how top spin correlations appear in the chain of processes 
 of $i\rightarrow t\bar{t}$ and decay of the top quarks.
The total matrix element squared 
 for the top-antitop pair production \p{top1} 
 and their decay channels \p{decay} is given by 
\begin{eqnarray}
|{\cal M}|^2 \propto {\rm Tr}[\rho R^i \bar{\rho}]
 =\rho_{\alpha^\prime\alpha}R^i_{\alpha\beta,\alpha^\prime\beta^\prime}
  \bar{\rho}_{\beta^\prime\beta} \label{comp}
\end{eqnarray}
in the narrow-width approximation for the top quark.
Here the subscripts denote the top and antitop spin indices, 
 and $R^i$ denotes the density matrix 
 corresponding to the production of the on-shell top-antitop quark pair 
 through the process $i$ in \p{top1}:
\begin{eqnarray}
 R_{\alpha\beta,\alpha^\prime\beta^\prime}^i
 =\sum_{{\rm initial~spin}}{\cal M}(i\rightarrow t_\alpha\bar{t}_\beta)
  {\cal M}^*(i\rightarrow t_{\alpha^\prime}\bar{t}_{\beta^\prime}),
\end{eqnarray}
where ${\cal M}(i\rightarrow t_\alpha\bar{t}_\beta)$ is 
 the amplitude for the top-antitop pair production. 
The matrices $\rho$ and $\bar{\rho}$ are the density matrices 
 corresponding to the decays of polarized top and antitop quarks 
 into some final states at the top and antitop rest frame, respectively.
In the leptonic decay modes, 
 the matrices $\rho$, which lead to \p{decay1},
 can be obtained as (see, for instance, \cite{Bernreuther})
\begin{eqnarray}
  \rho_{\alpha^\prime\alpha}
  = {\cal M}(t_\alpha \rightarrow bl^+\nu_l)
    {\cal M}^*(t_{\alpha^\prime} \rightarrow bl^+\nu_l)  
  = {\Gamma \over 2}(1 + \kappa_f {\vec{\sigma}} \cdot 
       \vec{q}_f)_{\alpha^\prime\alpha},  
 \label{rho1}
\end{eqnarray}
where $q_f$ is the unit vector of the direction of motion 
 of the decay product $f$. 
The density matrix for the polarized antitop quark 
 is obtained by replacing $\kappa_f \rightarrow -\kappa_f$ in \p{rho1}
 if there is no CP violation.
In the SM, there is no CP violation in the top quark decay at the
 leading order. 
In the model presented in the next section,
 there is no contribution to break CP symmetry at the leading order, and
 thus this relation holds.

A way to analyze the top-antitop spin correlations 
 is to see the angular correlations 
 of two charged leptons $l^+l^-$ 
 produced by the top-antitop quark leptonic decays. 
In the following, we consider only the leptonic decay channels. 
Using \p{comp}-\p{rho1} and integrating over 
 the azimuthal angles of the charged leptons, 
 we obtain the following double distribution
 \cite{Stelzer, Mahlon-Parke}
\begin{eqnarray}
 {1 \over \sigma}
 {d^2 \sigma \over d \cos\theta_{l^+} d \cos\theta_{l^-}}
 = {1 \over 4}\left({1+B_1 \cos\theta_{l^+}
 +B_2 \cos\theta_{l^-}-C
    \cos\theta_{l^+} \cos\theta_{l^-}}\right).
 \label{double}
\end{eqnarray}
Here $\sigma$ denotes the cross section 
 for the process of the leptonic decay modes, 
 and $\theta_{l^+} (\theta_{l^-})$ denotes the angle 
 between the top (antitop) spin axis and 
 the direction of motion of the antilepton (lepton) 
 at the top (antitop) rest frame. 
In what follows, we use the helicity spin basis which is almost optimal
 one to analyze the top spin correlation at the LHC\footnote{Recently another
 spin basis was constructed, which has a larger spin correlation than the
 helicity basis at the LHC \cite{uwer}.}. 
In this basis, the top (antitop)
 spin axis is regarded as the direction of motion of the top (antitop) in
 the top-antitop center-of-mass system.
The coefficients $B_1$ and $B_2$ are associated with a polarization of the top
 and antitop quarks, 
 and $C$ encodes the top spin correlations,
 whose explicit expression is given by
\begin{eqnarray}
 C= {\cal A} \kappa_{l^+}\kappa_{l^-},~~~\kappa_{l^+}=\kappa_{l^-}=1
\end{eqnarray}
where the coefficient ${\cal A}$ represents the spin asymmetry 
 between the produced top-antitop pairs 
 with like and unlike spin pairs defined as 
\begin{eqnarray}
 {\cal A}={\sigma(t_\uparrow\bar{t}_\uparrow)
          +\sigma(t_\downarrow\bar{t}_\downarrow) 
          -\sigma(t_\uparrow\bar{t}_\downarrow)
          -\sigma(t_\downarrow\bar{t}_\uparrow) 
         \over
           \sigma(t_\uparrow\bar{t}_\uparrow)
          +\sigma(t_\downarrow\bar{t}_\downarrow)
          +\sigma(t_\uparrow\bar{t}_\downarrow)
          +\sigma(t_\downarrow\bar{t}_\uparrow)}\,. 
\label{asym}
\end{eqnarray}
Here $\sigma(t_\alpha\bar{t}_\beta)$ is 
 the cross section of the top-antitop pair production 
 at parton level with denoted spin indices.

In the SM, at the lowest order of $\alpha_s$, 
 the spin asymmetry is found to be ${\cal A}=+0.319$ 
 for the LHC\footnote{
The parton distribution function set of CTEQ6L \cite{CTEQ} 
 has been used in our calculations. 
The resultant spin asymmetry somewhat depends 
 on the parton distribution functions used.}. 
At the LHC in the ATLAS experiment, the spin asymmetry 
 of the top-antitop pairs will be 
 measured with a precision of several percent, after one LHC year 
 at low luminosity (10 fb$^{-1}$) \cite{ATLAS_TOP}.

\section{A simple model with $Z'$ boson}
As a simple example which includes a $Z'$ boson, 
 we consider a model based on the gauge group
 $SU(3)\times SU(2)\times U(1)_1\times U(1)_2$ \cite{leike,Rizzo,Langacker}.
In order to realize the gauge symmetry breaking 
 $U(1)_1\times U(1)_2 \to U(1)_Y$, 
 where $U(1)_Y$ is the SM hypercharge group, 
 we introduce a scalar field $\Phi$ 
 in the representation $({\bf 1},{\bf 1}, +1,-1)$. 
The covariant derivative for the scalar field is given by 
\begin{eqnarray}
  {\cal D}_{\mu}^{\Phi} 
  = \partial_\mu - i g_1 B_{1\mu} + ig_2 B_{2\mu}\,,
\end{eqnarray}
where $g_i$ and $B_{i \mu}$ ($ i=1,2$) is the gauge coupling constant 
 and the gauge boson of $U(1)_i$. 
Once $\Phi$ develops a vacuum expectation value 
 ($ \langle \Phi \rangle =v_\Phi/\sqrt{2}$), 
 the gauge symmetry is broken down to 
 $U(1)_1\times U(1)_2 \to U(1)_Y$. 
Associated with this gauge symmetry breaking, 
 the mass eigenstates of two gauge bosons are described as  
\begin{eqnarray}
 B_\mu&=&B_{1\mu}\cos\phi+B_{2\mu}\sin\phi\,,\\
 Z_{2\mu}&=&-B_{1\mu}\sin\phi+B_{2\mu}\cos\phi\,,   
\end{eqnarray}
where $\phi$ is the mixing angle defined as $\tan\phi=g_1/g_2$, 
 $B_\mu$ is the massless $U(1)_Y$ gauge boson and 
 $Z_{2\mu}$ is the $Z^\prime$ boson with mass 
\begin{eqnarray}
 M_{Z_2}^2= 
  \left(\frac{g^\prime}{\sin\phi \cos\phi} \right)^2
   v_\Phi^2 \,.
\end{eqnarray} 
Here, the SM $U(1)_Y$ gauge coupling constant is defined 
 through the relation $1/{g}^{\prime 2}=1/g_1^2+1/g_2^2$.

For a fermion with a charge $(Y_{1f}, Y_{2f})$ 
 under $U(1)_1\times U(1)_2$, 
 the interaction term with $B_\mu$ and $Z_{2 \mu}$ is given by 
\begin{eqnarray}
 {\cal L}_{int} = 
 \bar{\psi}_f \gamma^\mu \psi_f \left[  
  g^\prime Y_{f} B_\mu + g^\prime 
  \left( -Y_{1f} \tan\phi + Y_{2f}\cot\phi \right) 
 Z_{2\mu}  \right] \, ,  
\end{eqnarray}
where $Y_f= Y_{1f} + Y_{2f}$. 
In our following analysis, we assume for simplicity that 
 under the gauge group, the quarks and leptons in each generation 
 have quantum numbers, 
 $q=({\bf 3}, {\bf 2},1/6,0)$, 
 $u_R=({\bf 3},{\bf 1}, 2/3,0)$, 
 $d_R=({\bf3},{\bf 1},-1/3,0)$, 
 $L=({\bf 1}, {\bf 2},-1/2,0)$, 
 $e_R=({\bf 1}, {\bf 1},-1,0)$ and 
 $\nu_R=({\bf 1}, {\bf 1}, 0, 0)$.

After the spontaneous breaking of the electroweak symmetry  
  $SU(2)\times U(1)_Y$ to the electromagnetic subgroup $U(1)_{EM}$, 
 the neutral current interactions for the SM leptons and quarks 
 are written as%
\footnote{
 Although the mixing between the SM $Z$ and $Z^\prime$ bosons, 
  in general, emerges through the electroweak symmetry breaking, 
  here we assume a negligibly small mixing, 
  which is required by the current precision measurements. 
 This situation is achievable in some models 
  even when the $Z^\prime$ boson mass is small 
  (see, for a concrete example, \cite{tophyper}). 
}
\begin{eqnarray}
 {\cal L}_{NC} = J_{EM}^\mu A_\mu
  +J_{Z_1}^\mu Z_{1\mu} + J_{Z_2}^\mu Z_{2\mu}\,.
\end{eqnarray}
Here $J_{EM}$ and $J_{Z_1}$ are the SM electromagnetic 
 and neutral currents 
\begin{eqnarray}
 J_{EM}^\mu=e\sum_f Q^f\bar{\psi}_f\gamma^\mu \psi_f\,,~~~
 J_Z^\mu&=&\sum_f \bar{\psi}_f\gamma^\mu(g_{L,1}^f P_L+g_{R,1}^f P_R)\psi_f\,,
\end{eqnarray}
where $e=g \sin\theta_W=g^\prime\cos\theta_W$,
 $Q^f$ is the electric charge of fermion $f$ and 
 $g_{L,1}^f$ and $g_{R,1}^f$ are the chiral couplings given by (see also
 Appendix \ref{coupling})
\begin{eqnarray}
 g_{L,1}^f={e \over \cos\theta_W\sin\theta_W}(T_{3L}^f-Q^f \sin^2\theta_W)\,,~~~
 g_{R,1}^f=-eQ^f\tan\theta_W\,
\end{eqnarray}
with the third component of weak isospin for the left chiral
 component of $\psi_f$, $T_{3L}^{f}$ ($T_{3L}^u=+1/2$ for up-type and 
 $T_{3L}^d=-1/2$ for down-type fermions). 
The current $J_{Z_2}$ is given as
\begin{eqnarray}
J_{Z_2}^\mu&=&g^\prime\sum_f (-Y_{1f}\tan\phi+Y_{2f}\cot\phi)\bar{\psi}_f \gamma^\mu \psi_f
 =\sum_f \bar{\psi}_f\gamma^\mu(g_{L,2}^f P_L+g_{R,2}^f P_R)\psi_f\,, 
\end{eqnarray}
where the chiral couplings $g_{L,2}^f$ and $g_{R,2}^f$ are explicitly 
 given in Appendix \ref{coupling}. 
In the following analysis, 
 we fix the mixing angle $\phi$ as $\tan\phi=1$, 
 for simplicity.

\section{Scattering amplitude}
In this section we calculate the squared invariant amplitudes for
 $q\bar{q}\rightarrow t\bar{t}$ and $gg\rightarrow t\bar{t}$ processes.
First we calculate $q\bar{q}\rightarrow  t\bar{t}$ process.
In this process, top quark pairs are produced via the $s$-channel
 photon, $Z$ boson and $Z^\prime$ boson exchanges.
An amplitude for quark annihilation process is given by 
\begin{eqnarray}
 {\cal M}(q\bar{q}\rightarrow t\bar{t})&=&{\cal M}_{\rm
  QCD}+{\cal M}_{\rm NC}\,.
\end{eqnarray}
Here ${\cal M}_{\rm QCD}$ denotes the QCD process and
 ${\cal M}_{\rm NC}$ is the contribution of the neutral current.
Since there is no interference between the QCD process
 and the neutral current process, the squared amplitude is simply
 written as 
\begin{eqnarray}
 |{\cal M}|^2=|{\cal M}_{\rm QCD}|^2+|{\cal M}_{\rm NC}|^2\,.
\end{eqnarray}
The helicity amplitude of the QCD process, ${\cal M}_{\rm QCD}$, 
 is given by (for one flavor initial state)
\begin{eqnarray}
 |{\cal M}(q\bar{q}\rightarrow t_\uparrow\bar{t}_\uparrow)|^2
  &=& |{\cal M}(q\bar{q}\rightarrow {t_\downarrow\bar{t}_\downarrow})|^2
  = {g_s^4 \over 9}(1-\beta_t^2)\sin^2\theta\,,
  \label{qq1} \\
 |{\cal M}(q\bar{q}\rightarrow t_\uparrow\bar{t}_\downarrow)|^2
  &=& |{\cal M}(q\bar{q}\rightarrow {t_\downarrow\bar{t}_\uparrow})|^2
  = {g_s^4 \over 9}(1+\cos^2\theta)\,,
  \label{qq2}
\end{eqnarray}
where $q(\bar{q})$ denotes an initial (anti)quark, $g_s$ is the strong
coupling constant, $\beta_t=\sqrt{1-4m_t^2/s}$, $m_t$ is the top quark
 mass, $\theta$ is the scattering angle between incoming quark and outgoing
 top quark, and $\sqrt{s}$ is energy of colliding partons.
The helicity amplitude of the neutral current process, 
 ${\cal M}_{\rm NC}(q\bar{q}\rightarrow t\bar{t})$, 
 is described by 
\begin{eqnarray}
|{\cal M}_{\rm NC}(q\bar{q}\rightarrow t_\delta\bar{t}_\gamma)|^2=\left({1
 \over 2}\right)^2
 \sum_{\alpha,\beta}|{\cal M}_{\rm NC}(\alpha,\beta;\delta,\gamma)|^2\,,
\end{eqnarray}
where ${\cal M}_{\rm NC}(\alpha,\beta;\delta,\gamma)$ is the decomposition
 of the helicity amplitude with respect to the initial spin
 and $\alpha(\delta)$ and $\beta(\gamma)$ denote initial (final) spin states
 for quark and antiquark, respectively.
The form is described by
\begin{eqnarray}
&&\hspace{-10mm}{\cal M}_{\rm NC}(+,-;\pm,\pm)=\mp s\sqrt{1-\beta_t^2}\sin\theta
  \left[{(eQ^f)(eQ^t) \over s}+
  \sum_{i=1}^2{g_{R,i}^f \over 2}{g_{L,i}^t+g_{R,i}^t \over
  s-M_{Z_i}^2+iM_{Z_i}\Gamma_{Z_i}} \right]\,,\\
&&\hspace{-10mm}{\cal M}_{\rm NC}(-,+;\pm,\pm)=\mp s\sqrt{1-\beta_t^2}\sin\theta
  \left[{(eQ^f)(eQ^t) \over s}+
  \sum_{i=1}^2{g_{L,i}^f \over 2}{g_{L,i}^t+g_{R,i}^t \over
  s-M_{Z_i}^2+iM_{Z_i}\Gamma_{Z_i}} \right]\,, \\
&&\hspace{-10mm}{\cal M}_{\rm NC}(+,-;+,-)=-s(1+\cos\theta)\left[
  {(eQ^f)(eQ^t) \over s}+\sum_{i=1}^2{g_{R,i}^f \over 2}
 {(1-\beta_t)g_{L,i}^t+(1+\beta_t)g_{R,i}^t \over s-M_{Z_i}^2+iM_{Z_i}\Gamma_{Z_i}}
  \right]\,,\\
&&\hspace{-10mm}{\cal M}_{\rm NC}(+,-;-,+)=s(1-\cos\theta)\left[
  {(eQ^f)(eQ^t)\over s}+\sum_{i=1}^2{g_{R,i}^f \over 2}
 {(1+\beta_t)g_{L,i}^t+(1-\beta_t)g_{R,i}^t \over s-M_{Z_i}^2+iM_{Z_i}\Gamma_{Z_i}}
  \right]\,,\\
&&\hspace{-10mm}{\cal M}_{\rm NC}(-,+;+,-)=s(1-\cos\theta)\left[
  {(eQ^f)(eQ^t) \over s}+\sum_{i=1}^2{g_{L,i}^f \over 2}
 {(1-\beta_t)g_{L,i}^t+(1+\beta_t)g_{R,i}^t \over s-M_{Z_i}^2+iM_{Z_i}\Gamma_{Z_i}}
  \right]\,,\\
&&\hspace{-10mm}{\cal M}_{\rm NC}(-,+;-,+)=-s(1+\cos\theta)\left[
  {(eQ^f)(eQ^t) \over s}+\sum_{i=1}^2{g_{L,i}^f \over 2}
 {(1+\beta_t)g_{L,i}^t+(1-\beta_t)g_{R,i}^t \over s-M_{Z_i}^2+iM_{Z_i}\Gamma_{Z_i}}
  \right]\,, \label{gamma}
\end{eqnarray}
with the decay widths of $Z_1$ and $Z_2$ bosons given by
\begin{eqnarray}
 \Gamma_{Z_i}=\Gamma(Z_i\rightarrow f\bar{f})
 ={M_{Z_i}\over 96\pi}\sum_{f}\beta_i^f
  \left\{(3+(\beta_i^{f})^2)((g_{L,i}^f)^2+(g_{R,i}^f)^2)+6(1-(\beta_i^f)^2)g_{L,i}^f
   g_{R,i}^f\right\}\,,\label{gamma-total}
\end{eqnarray}
where $M_{Z_i}$ is the mass of the gauge boson $Z_i$,
 $\beta_i^f=\sqrt{1-4m_f^2/M_{Z_i}^2}$,
 $m_f$ is the mass of the fermion $f$. 
The explicit form of $\Gamma_{Z_i}$ for each decay mode is given in
 Appendix \ref{appendix-C}.
The sum in Eq. (\ref{gamma-total}) does not include 
 the right-handed neutrinos since their masses are naturally around $v_R$.

For the squared amplitude for the QCD process with the $gg$ initial state, we have
\begin{eqnarray} 
 &&|{\cal M}(gg \rightarrow t_\uparrow\bar{t}_\uparrow)|^2
    = |{\cal M}(gg \rightarrow {t_\downarrow\bar{t}_\downarrow})|^2
 ={g_s^4 \over 96}{\cal Y}(\beta_t,\cos\theta)(1-\beta_t^2)
     (1+\beta_t^2+\beta_t^2\sin^4\theta)\,,\label{gg1}\\
 &&|{\cal M}(gg\rightarrow t_\uparrow\bar{t}_\downarrow)|^2
   = |{\cal M}gg\rightarrow {t_\downarrow\bar{t}_\uparrow})|^2
   ={g_s^4 \beta_t^2 \over 96}{\cal
     Y}(\beta_t,\cos\theta)\sin^2\theta(1+\cos^2\theta)\,. \label{gg2}
\end{eqnarray}
Here $ {\cal Y}(\beta_t,\theta)$ is defined by
\begin{eqnarray}
 {\cal Y}(\beta_t,\cos\theta)&=&
  {7+9\beta_t^2\cos^2 \theta \over (1-\beta_t^2\cos^2\theta)^2}\,.
  \label{funcs1}
\end{eqnarray}

Using above formulas, we calculate the double distribution
 (\ref{double}) including the $Z'$ contributions.
Explicit calculation tells us that the transverse polarization is
 vanishing, i.e. $B_1=B_2=0$ while the spin asymmetry ${\cal A}$ 
 is altered from the SM one.
\section{Numerical results}
Here we show various numerical results 
 and demonstrate interesting properties of measurable quantities in our model. 
In our analysis we use the parton distribution function of 
 CTEQ6L \cite{CTEQ} with the factorization scale $Q=m_t=175$
 GeV and $\alpha_s(Q)=0.1074$. 
We choose $M_{Z_2}=900$ GeV to be consistent with the current experimental 
 results \cite{CDF}.
In the whole analysis, the center of mass energy of the colliding
 protons, $E_{CMS}$, is taken to be $14$ TeV at the LHC.

\begin{figure}[ht]
\begin{center}
  \epsfxsize=8.2cm
  \epsfbox{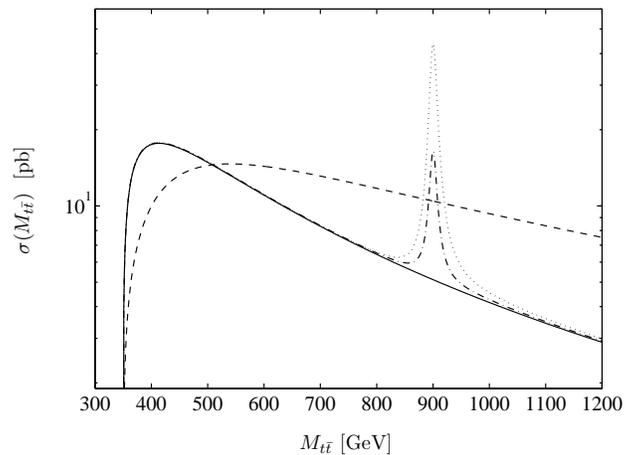}
\caption{The dependence of the  
 cross section of the top-antitop 
 quark pair production by quark pair annihilation and gluon fusion
 on the center-of-mass energy of colliding partons $M_{t\bar{t}}$. 
The solid and dashed lines correspond to 
 the results of the quark annihilation and gluon fusion for the SM, respectively.
The dotted and dash-dotted lines correspond to the results of
 $u\bar{u}\rightarrow t\bar{t}$ and $d\bar{d}\rightarrow t\bar{t}$, respectively.}
  \label{cross}
\end{center}
\end{figure}

Fig. \ref{cross} shows the cross sections of the top-antitop pair
 production through $u\bar{u} \rightarrow t\bar{t}$ 
 and $d\bar{d} \rightarrow t\bar{t}$ at the parton level 
 as a function of parton center-of-mass energy
 $\sqrt{s}=M_{t\bar{t}}=\sqrt{(p_t+p_{\bar{t}})^2}$, 
 where $p_t(p_{\bar{t}})$ is a momentum of (anti)top quark.
The figure exhibits the large peak corresponding 
 to the resonant production of $Z^\prime$ boson.

\begin{figure}[ht]
\begin{center}
  \epsfxsize=8.2cm
  \epsfbox{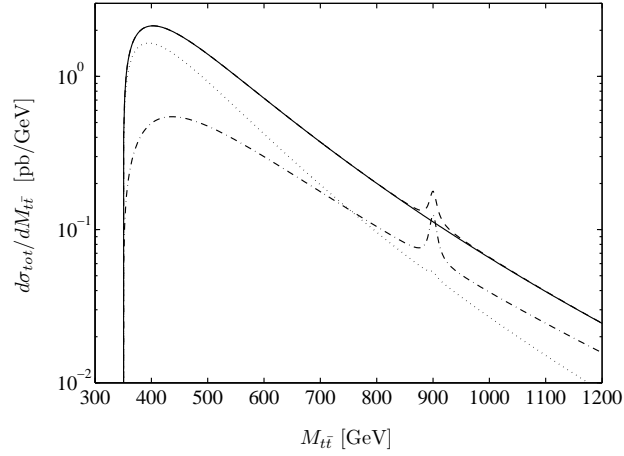}
\caption{Differential cross section \p{total_s} as a function 
 of the top-antitop invariant mass $M_{t\bar{t}}$. 
The solid and dashed lines correspond to 
 the results of the SM and our model.
The breakdown of the differential cross sections into the like (dotted) and the unlike
 (dash-dotted) top-antitop spin pair productions are also depicted.}
  \label{cross-inv}
\end{center}
\end{figure}

The dependence of the cross section 
 on the top-antitop invariant mass $M_{t\bar{t}}$
 is given by 
\begin{eqnarray}
 \frac{d \sigma_{tot}(pp \rightarrow t\bar{t})}{d\sqrt{s}}=
  \sum_{a,b} \int\limits_{-1}^{1} d \cos\theta
   \int\limits_{\frac{s}{E_{CMS}^2}}^1 d x_1 
   \frac{2\sqrt{s}}{x_1 E_{CMS}^2} f_a(x_1,Q^2)
   f_b\left(\frac{s}{x_1 E_{CMS}^2},Q^2\right)
   {d\sigma(t\bar{t}) \over d\cos \theta}.
   \label{total_s}
\end{eqnarray}
Fig. \ref{cross-inv} depicts the result of the differential cross
 section (\ref{total_s}).
Here, the decomposition of the total cross section into the 
 like ($t_\uparrow\bar{t}_\uparrow+t_\downarrow\bar{t}_\downarrow$) 
 and unlike ($t_\uparrow\bar{t}_\downarrow+t_\downarrow\bar{t}_\uparrow$)  
 top-antitop spin pairs is also shown.
The deviation from the SM around the resonant pole is large only 
 for unlike top-antitop
 spin pairs, because of the helicity conserving interactions
 between the $Z^\prime$ boson and the SM fermions.
Note that the resonance of $Z^\prime$ will be firstly measured in the
 leptonic process $pp\rightarrow \mu^+\mu^-$ since the background is
 quite small. 
The information of the pole will be then confirmed in the top-antitop
 quark production.
Furthermore it should be useful to observe large deviation of the top
 spin correlation as we will discuss below.

\begin{figure}[htb]
\begin{center}
  \epsfxsize=8.2cm
  \epsfbox{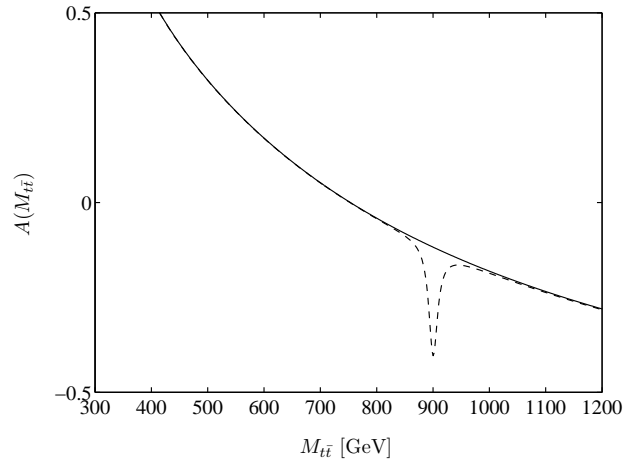}
\caption{
Spin asymmetry $\cal{A}$ as a function of 
 the top-antitop invariant mass $M_{t\bar{t}}$. 
The solid line corresponds to the SM, 
 while the dashed line corresponds to our model.}
  \label{spin1}
\end{center}
\end{figure}

\begin{figure}[htb]
\begin{center}
  \epsfxsize=8.2cm
  \epsfbox{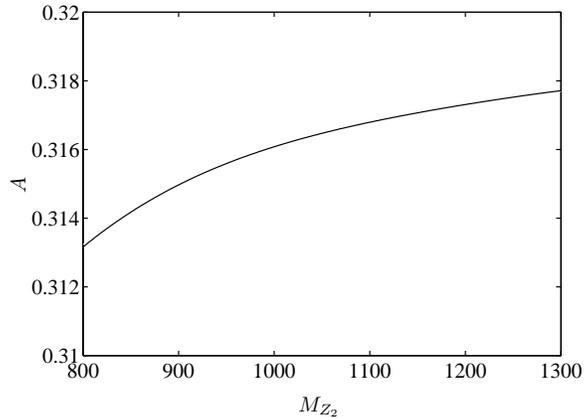}
\caption{
Spin asymmetry $\cal{A}$ as a function of 
 $M_{Z_2}$.}
  \label{spin2}
\end{center}
\end{figure}

Now we show the result for the spin asymmetry ${\cal A}$ in
 Fig. \ref{spin1} as a function of the
 center-of-mass energy of colliding partons.
As expected, deviation from the SM is enhanced around the $Z^\prime$
 boson resonance pole.
In Fig.~\ref{spin2}, we show the spin asymmetry 
 $\cal{A}$ as a function of the $Z'$ boson mass $M_{Z_2}$,
 after integration with respect to $M_{t\bar{t}}$ in the range
 $2m_t\le M_{t\bar{t}}\le E_{CMS}$.
For $M_{Z_2}\gtrsim 900$ GeV,  deviation from the SM is less than
 a few percent. 
In order to enhance the deviation of the spin asymmetry from the SM one,
 we impose the selection criteria
 $M_{Z_2}-50\ {\rm GeV} < M_{t\bar{t}} < M_{Z_2}+50\ {\rm GeV}$ for the
 range of the integration.
Table \ref{table} presents values of the spin asymmetry for
 three chosen masses $M_{Z_2}=$ 900, 1100 and 1300 GeV. 
We see sizable deviations from the SM ones. 
The corresponding values of the total cross section\footnote{
Cross sections in Table \ref{table} are computed in the leading order.
 In the SM, NLO contributions significantly increase the total cross section,
 nevertheless the spin asymmetry is insensitive to NLO contributions \cite{Bernreuther2}.} 
 imply statistically sufficient number of events in the 
 selected kinematical range
 (under assumption of integral luminosity ${\cal L}=10$ fb$^{-1}$ 
  for one LHC year at low luminosity and ${\cal L}=100$ fb$^{-1}$ 
  for one LHC year at high luminosity).

\section{Conclusions} 
We studied top-antitop pair production and its spin correlation in the
 $Z'$ model at the LHC. 
In this model, in addition to the SM processes,
 there is a new contribution to the top-antitop
 pair production process mediated by an electrically neutral gauge
 boson $Z'$ in the $s$-channel.
We calculated the squared invariant amplitudes for the top-antitop pair
 production including the new contribution from the $Z^\prime$ boson and
 showed numerical results for the top
 pair production cross section and the top spin correlations.
We found that after integration with respect to $M_{t\bar{t}}$ for the
 full range $2m_t\le M_{t\bar{t}}\le E_{CMS}$, the deviation of the spin
 asymmetry $\cal{A}$ from the SM result is very small, below the
 estimated sensitivity of the ATLAS experiment \cite{ATLAS_TOP}. 
When we imposed the selection criteria 
 $M_{Z_2}-50\ {\rm GeV} < M_{t\bar{t}}< M_{Z_2}+50\ {\rm GeV}$ for the 
 range of the integration, the deviation of the spin asymmetry from the
 SM one is remarkably enhanced (around $50\%$ of the SM value), even if the total
 cross sections are almost the same. 
By using the same analysis performed in \cite{ATLAS_TOP}, we can estimate the
 sensitivity for our model.
In Ref. \cite{ATLAS_TOP}, it is shown that the spin asymmetry of 
 the top-antitop pairs in the SM will be measured with a precision of 
 $6\%$ after one LHC year at low luminosity, $10$ fb${}^{-1}$. 
In our model,
 for instance, for the case of $M_{Z_2}=900$ GeV and
 $\sigma^{(cut)}=12.8$ pb, 
 we will reach the same accuracy of the measurement of the spin
 correlation with that
 of the SM when the integrated luminosity is $270$ fb${}^{-1}$ 
 ($= 2.7$ years of high luminosity LHC run).
Note that it is very rough estimation since the sensitivity of the 
 ATLAS experiment on
 the spin correlation published in \cite{ATLAS_TOP} was estimated selecting low
 energetic top quarks with $M_{t\bar{t}} < 550$ GeV.
In order to estimate the sensitivity more accurately with a high
 $M_{t\bar{t}}$ region for our case, we need Monte-Carlo simulations
 including the detector response.
It must be the subject of next study. 

Finally we compare our results with our previous analysis for
 Kaluza-Klein productions in the large extra-dimensions \cite{AOSS1} 
 and the Randall-Sundrum model \cite{AOSS2}.
All these models affect to reduce the spin 
 asymmetry ${\cal A}$ from the SM results,
 when new physics effects are sizable.
This is because the interactions between new particles and top quarks
 conserve chirality and as a result the cross section of the
 top-antitop quark pair production for the like spin pair is enhanced from
 the SM one.
In our model with $Z'$ boson, the situation is the same and the spin
 asymmetry is decreased. 

In summary, the top spin correlations is a powerful tool to reveal the property of
 the $Z'$ boson at the LHC.

{\bf Note added}:
Recently top quark pair productions and their spin correlations 
 through possible new physics $s$-channel resonances have been 
 investigated in a model independent way with the use of 
 Monte-Carlo simulation \cite{maltoni}. 
In their analysis, a $Z'$ gauge boson couples with the same strength to
 fermions as the SM $Z$ boson.
Then the peak of the $Z'$ resonance pole in the cross section
 against $M_{t\bar{t}}$ is larger than ours for 
 the same $Z'$ boson masses.
Therefore the absolute value of the peak corresponding 
 the $Z'$ boson resonance in the top spin correlation against 
 $M_{t\bar{t}}$ is larger than ours and consequently
 their spin asymmetry is much smaller than ours.
We have checked that our results are consistent 
 with ones in \cite{maltoni} 
 when we choose the same setting as in \cite{maltoni}.
\\\\
%
%
\noindent{\Large \bf Acknowledgements}

The work of M.A. is supported by the Science Research Center Program of
 the Korea Science and Engineering Foundation through the Center for
 Quantum Spacetime (CQUeST) of Sogang University with grant number
 R11-2005-021.
M.A. would also like to thank to Czech Technical University in Prague
 and theory division, KEK, for their 
 hospitality during his visit.
The work of N.O. is supported in part by Scientific Grants 
 from the Ministry of Education, and Science and Culture of Japan
 (No. 18740170) and the Academy of Finland Finnish-Japanese Core Programme grant 112420.
N.O. would also like to thank the High Energy Physics Division of the Department
 of Physical Sciences, University of Helsinki, for their hospitality during his visit.
The work of K.S. is supported by the Research Program  
 MSM6840770029 and by the project of International Cooperation ATLAS-CERN 
 of the Ministry of Education, 
 Youth and Sports of the Czech Republic.

\newpage
\noindent{\Large \bf Appendix}
\appendix
\section{General formula}\label{hel-amp}
Consider the following Lagrangian
\begin{eqnarray}
 {\cal L}_{\rm int}^f=eJ^\mu Z_{\mu}
 =\bar{\psi}_f \gamma^\mu(g_L^f P_L+g_R^fP_R)\psi_f Z_{\mu}\,,
\end{eqnarray}
where $Z_\mu$ denotes a massive gauge boson whose mass is $M$.
A polarized invariant amplitude for the process 
 $f(\alpha)\bar{f}(\beta)\rightarrow F(\delta)\bar{F}(\gamma)$ is given by
\begin{eqnarray}
 {\cal M}(\alpha,\beta;\gamma,\delta)={1 \over
  p^2 - M^2 + iM\Gamma_{Z}}J_{\rm in}^\mu(\alpha,\beta)
  J_{{\rm out}\mu}(\gamma,\delta)\,,
\end{eqnarray}
where $\alpha,\beta$($\gamma,\delta$) denote initial(final) spin states
 for quark and anti-quark, respectively, and $\Gamma$ is the decay width
 of $Z$ boson.
The currents of gauge bosons for initial (massless) and final (massive) states are given by
\begin{eqnarray}
 J_{\rm in}^\mu(+,-)=-\sqrt{s}g_R^f(0,1,i,0)\,,~~~~~
 J_{\rm in}^\mu(-,+)=-\sqrt{s}g_L^f(0,1,-i,0)\,,
\end{eqnarray}
and
\begin{eqnarray}
 J_{\rm
  out}^{\mu}(+,+)&=&\omega_+\omega_-\left\{g_L^F(1,-\sin\theta,0,-\cos\theta)
  -g_R^F(1,\sin\theta,0,\cos\theta)\right\}\,,\\
 J_{\rm
  out}^{\mu}(-,-)&=&\omega_+\omega_-\left\{g_L^F(1,\sin\theta,0,\cos\theta)
  -g_R^F(1,-\sin\theta,0,-\cos\theta)\right\}\,,\\
 J_{\rm
  out}^{\mu}(+,-)&=&\omega_-^2g_L^F(0,-\cos\theta,i,\sin\theta)
  -\omega_+^2g_R^F(0,\cos\theta,-i,-\sin\theta)\,,\\
 J_{\rm
  out}^{\mu}(-,+)&=&\omega_+^2g_L^F(0,-\cos\theta,-i,\sin\theta)
  -\omega_-^2g_R^F(0,\cos\theta,i,-\sin\theta)\,,
\end{eqnarray}
where $\omega_\pm^2={\sqrt{s}\over 2}(1\pm \beta_F)$, 
 $\beta_F=\sqrt{1-{4m_F^2 \over s}}$ and 
 $f$($F$) denotes a flavor of initial (final) state of fermions.

\section{Couplings}\label{coupling}
The couplings for the SM $Z$ boson:
\begin{eqnarray}
 g_{L,1}^\nu&=&{e \over \cos\theta_W\sin\theta_W}{1 \over
  2}\,,~~~g_{R,1}^\nu=0\,,\\
 g_{L,1}^l&=&{e \over \cos\theta_W\sin\theta_W}
   \left(-{1 \over 2}-\sin^2\theta_W(-1)\right)\,,
  ~~~g_{R,1}^l=-e(-1)\tan\theta_W\,,\\
 g_{L,1}^u&=&{e \over \cos\theta_W\sin\theta_W}
   \left({1 \over 2}-\sin^2\theta_W{2 \over 3}\right)\,,
  ~~~g_{R,1}^u=-e{2 \over 3}\tan\theta_W\,,\\
 g_{L,1}^d&=&{e \over \cos\theta_W\sin\theta_W}
   \left(-{1 \over 2}-\sin^2\theta_W\left(-{1 \over 3}\right)\right)\,,
  ~~~g_{R,1}^d=-e\left(-{1 \over 3}\right)\tan\theta_W\,,
\end{eqnarray}
The couplings for $Z^\prime$ boson: 
\begin{eqnarray}
 g_{L,2}^\nu&=&{e \over \cos\theta_W}{1 \over  2}\tan\phi\,,~~~g_{R,2}^\nu=0\,,\\
 g_{L,2}^l&=&{e \over \cos\theta_W}{1 \over  2}\tan\phi\,,
  ~~~g_{R,2}^l={e \over \cos\theta_W}\tan\phi\,,\\
 g_{L,2}^u&=&{e \over \cos\theta_W}\left(-{1 \over  6}\right)\tan\phi\,,
  ~~~g_{R,2}^u={e \over \cos\theta_W}\left(-{2 \over 3}\right)\tan\phi\,,\\
 g_{L,2}^d&=&{e \over \cos\theta_W}\left(-{1 \over  6}\right)\tan\phi\,,
  ~~~g_{R,2}^d={e \over \cos\theta_W}{1 \over 3}\tan\phi\,.
\end{eqnarray}
\section{Decay width}\label{appendix-C}
The decay widths of $Z(Z_1)$ and $Z^\prime(Z_2)$ bosons:
\begin{eqnarray}
 \Gamma(Z_i\rightarrow \nu\bar{\nu})&=&{M_{Z_i} \over
  24\pi}((g_{L,i}^{\nu})^2+(g_{R,i}^{\nu})^2)\,,\\
 \Gamma(Z_i\rightarrow l\bar{l})&=&{M_{Z_i} \over
  24\pi}((g_{L,i}^{l})^2+(g_{R,i}^l)^2)\,,\\
 \Gamma(Z_i\rightarrow u\bar{u})&=&{M_{Z_i} \over
  24\pi}3((g_{L,i}^{u})^2+(g_{R,i}^u)^2)\,,\\
 \Gamma(Z_i\rightarrow d\bar{d})&=&{M_{Z_i} \over
  24\pi}3((g_{L,i}^d)^2+(g_{R,i}^d)^2)\,,\\
 \Gamma(Z_i\rightarrow t\bar{t})&=&{M_{Z_i} \over
  24\pi}3\beta_t\left[\left(1-{m_t^2 \over
 M_{Z_i}^2}\right)((g_{L,2}^t)^2+(g_{R,2}^t)^2)+6{m_t^2
  \over M_{Z_i}^2}g_{L,2}^tg_{R,2}^t\right]\delta_{i2}\,,
\end{eqnarray}
where $\beta_t=\sqrt{1-4m_t^2/M_{Z_2}^2}$.
%
%
%

\newpage
%
%
%
\begin{table}[b]
\begin{center}
\begin{tabular}{c c c c c c c}
\hline
$M_{Z_2}$ [GeV] & ${\cal A}$ & $\sigma$ [pb] & ${\cal A}^{({\rm cut})}$ & ${\cal A}_{\rm SM}^{({\rm cut})}$ 
 & $\sigma^{({\rm cut})}$ [pb] & $\sigma_{SM}^{({\rm cut})}$ [pb]\\
\hline
 900 & 0.315  & 491 & $-0.199$ & $-0.114$ & 12.8 & 11.4 \\
1100 & 0.317  & 490 & $-0.331$ & $-0.232$ & 4.65 & 3.99 \\
1300 & 0.317  & 490 & $-0.430$ & $-0.319$ & 1.91 & 1.57 \\
\hline
SM   & 0.319  & 489 &          &          &    &     \\
\hline
\end{tabular}
\end{center}
\caption{
Spin asymmetry ${\cal A}$ and $t\bar{t}$ total cross section
 for the top-antitop events without the constraint on the invariant mass
 (second and third column) and with the invariant mass cut in the range 
 $M_{Z_2}-50\ {\rm GeV} < M_{t\bar{t}} < M_{Z_2}+50\ {\rm GeV}$ 
 (from fourth to seventh column) for various values of $M_{Z_2}$.
The values in the fifth and seventh column correspond to the SM,
 while in the other columns correspond to our $Z'$ model.
The last line shows the SM results.
 }
\label{table}
\end{table}

\begin{thebibliography}{100}
%
\bibitem{LR}
  J.~C.~Pati and A.~Salam,
  Phys.\ Rev.\  D {\bf 10} (1974) 275
  [Erratum-ibid.\  D {\bf 11} (1975) 703];
%
  R.~N.~Mohapatra and J.~C.~Pati,
  Phys.\ Rev.\  D {\bf 11} (1975) 566, 2559;
  G.~Senjanovic and R.~N.~Mohapatra,
  Phys.\ Rev.\  D {\bf 12} (1975) 1502;
  G.~Senjanovic,
  Nucl.\ Phys.\  B {\bf 153} (1979) 334.
%
\bibitem{SO10}
 R.~N.~Mohapatra, {\it Unification and  Supersymmetry}, Springer, New
	York, 1986.
%
\bibitem{E6}
  F.~del Aguila, M.~Quiros and F.~Zwirner,
  Nucl.\ Phys.\  B {\bf 284} (1987) 530;
%
  J.~L.~Hewett and T.~G.~Rizzo,
  Phys.\ Rept.\  {\bf 183} (1989) 193.
%
\bibitem{leike}
  A.~Leike,
  Phys.\ Rept.\  {\bf 317} (1999) 143
  [arXiv:hep-ph/9805494].
%
\bibitem{Rizzo}
  T.~G.~Rizzo,
  ``Z' phenomenology and the LHC,''
  arXiv:hep-ph/0610104.
%
\bibitem{Langacker}
  P.~Langacker,
  ``The Physics of Heavy Z' Gauge Bosons,''
  arXiv:0801.1345 [hep-ph].
%
\bibitem{CDF}
  A.~Abulencia {\it et al.}  [CDF Collaboration],
  Phys.\ Rev.\ Lett.\  {\bf 96} (2006) 211801
  [arXiv:hep-ex/0602045].
%
\bibitem{fuks}
  B.~Fuks, M.~Klasen, F.~Ledroit, Q.~Li and J.~Morel,
  ``Precision predictions for Z'-production at the CERN LHC: QCD matrix
  elements, parton showers, and joint resummation,''
  arXiv:0711.0749 [hep-ph].
%
\bibitem{petriello}
  F.~Petriello and S.~Quackenbush,
  ``Measuring Z' couplings at the LHC,''
  arXiv:0801.4389 [hep-ph].
%
\bibitem{guzzi}
  C.~Coriano, A.~E.~Faraggi and M.~Guzzi,
  ``Searching for Extra Z' from Strings and Other Models at the LHC with
  Leptoproduction,''
  arXiv:0802.1792 [hep-ph].
%
\bibitem{Abe}
  F.~Abe {\it et al.}  [CDF Collaboration],
  Phys.\ Rev.\ Lett.\  {\bf 74}, 2626 (1995)
  [arXiv:hep-ex/9503002].
%
\bibitem{Bigi}
  I.~I.~Y.~Bigi, Y.~L.~Dokshitzer, V.~A.~Khoze, J.~H.~K\"uhn and P.~M.~Zerwas,
  Phys.\ Lett.\ B {\bf 181}, 157 (1986).
%
\bibitem{Stelzer}
  T.~Stelzer and S.~Willenbrock,
  Phys.\ Lett.\ B {\bf 374}, 169 (1996)
  [arXiv:hep-ph/9512292];
  A.~Brandenburg,
  Phys.\ Lett.\ B {\bf 388}, 626 (1996)
  [arXiv:hep-ph/9603333];
  D.~Chang, S.~C.~Lee and A.~Sumarokov,
  Phys.\ Rev.\ Lett.\  {\bf 77}, 1218 (1996)
  [arXiv:hep-ph/9512417].
\bibitem{Mahlon-Parke}
  G.~Mahlon and S.~J.~Parke,
  Phys.\ Rev.\ D {\bf 53}, 4886 (1996)
  [arXiv:hep-ph/9512264];
  Phys.\ Lett.\ B {\bf 411}, 173 (1997)
  [arXiv:hep-ph/9706304].
%
\bibitem{Bernreuther2}
  W.~Bernreuther, A.~Brandenburg, Z.~G.~Si and P.~Uwer,
  Phys.\ Rev.\ Lett.\  {\bf 87}, 242002 (2001)
  [arXiv:hep-ph/0107086];
  Nucl.\ Phys.\ B {\bf 690}, 81 (2004)
  [arXiv:hep-ph/0403035].
%
\bibitem{AOSS1}
  M.~Arai, N.~Okada, K.~Smolek and V.~\v{S}im\'ak,
  Phys.\ Rev.\ D {\bf 70} (2004) 115015
  [arXiv:hep-ph/0409273].
%
\bibitem{AOSS2}
  M.~Arai, N.~Okada, K.~Smolek and V.~Simak,
  Phys.\ Rev.\  D {\bf 75} (2007) 095008
  [arXiv:hep-ph/0701155].
%
\bibitem{ee}
  K.~Y.~Lee, H.~S.~Song, J.~H.~Song and C.~Yu,
  Phys.\ Rev.\  D {\bf 60} (1999) 093002
  [arXiv:hep-ph/9905227];
%
  K.~Y.~Lee, S.~C.~Park, H.~S.~Song and C.~Yu,
  Phys.\ Rev.\  D {\bf 63} (2001) 094010
  [arXiv:hep-ph/0011173];
%
  C.~X.~Yue, L.~Wang, L.~N.~Wang and Y.~M.~Zhang,
  Chin.\ Phys.\ Lett.\  {\bf 23} (2006) 2379;
%
  B.~Sahin,
  arXiv:0802.1937 [hep-ph].
%
\bibitem{gamma}
 K.~Y.~Lee, S.~C.~Park, H.~S.~Song, J.~H.~Song and C.~Yu,
  Phys.\ Rev.\  D {\bf 61} (2000) 074005
  [arXiv:hep-ph/9910466];
%
  I.~Sahin,
  arXiv:0802.2818 [hep-ph].
%
\bibitem{Jezabek}
  A.~Czarnecki, M.~Jezabek and J.~H.~K\"uhn,
  Nucl.\ Phys.\ B {\bf 351} (1991) 70.
%
\bibitem{Bernreuther}
  W.~Bernreuther, O.~Nachtmann, P.~Overmann and T.~Schr\"oder,
  Nucl.\ Phys.\ B {\bf 388}, 53 (1992)
  [Erratum-ibid.\ B {\bf 406}, 516 (1993)];
  A.~Brandenburg and J.~P.~Ma,
  Phys.\ Lett.\ B {\bf 298}, 211 (1993).
%
\bibitem{uwer}
  P.~Uwer,
  Phys.\ Lett.\  B {\bf 609} (2005) 271
  [arXiv:hep-ph/0412097].
%
\bibitem{CTEQ}
  J.~Pumplin, D.~R.~Stump, J.~Huston, H.~L.~Lai, P.~Nadolsky and W.~K.~Tung,
  JHEP {\bf 07} (2002) 012
  [arXiv:hep-ph/0201195].
%
\bibitem{ATLAS_TOP}
  F.~Hubaut, E.~Monnier, P.~Pralavorio, V.~\v{S}im\'ak, K.~Smolek,
  Eur.\ Phys.\ J.\ C {\bf 44} (2005) 13
  [arXiv:hep-ex/0508061].
%
\bibitem{tophyper} 
  C.~W.~Chiang, J.~Jiang, T.~Li and Y.~R.~Wang,
  JHEP {\bf 0712}, 001 (2007)
  [arXiv:0710.1268 [hep-ph]].
%
\bibitem{maltoni}
  R.~Frederix and F.~Maltoni,
  ``Top pair invariant mass distribution: a window on new physics,''
  arXiv:0712.2355 [hep-ph].
%

\end{thebibliography}
\end{document}